\RequirePackage[final]{graphicx}
\documentclass[aps, superscriptaddress,prr,twocolumn,footinbib]{revtex4-1}
\usepackage[utf8]{inputenc}
\usepackage{amsmath,amsfonts,amsthm,bm}
\usepackage{mathtools}
\usepackage{graphicx}
\usepackage{braket}
\usepackage{cleveref}
\usepackage{fixme}
\usepackage{color}

\newcommand{\nn}{\nonumber}

\renewcommand{\braket}[1]{\left\langle #1 \right\rangle}

\renewcommand{\Im}[0]{\text{Im}}

\renewcommand{\k}[0]{\mathbf{k}}
\newcommand{\q}[0]{\mathbf{q}}
\renewcommand{\r}[0]{\mathbf{r}}

\renewcommand{\eqref}[1]{(\ref{#1})}
\renewcommand{\vec}[1]{\mathbf{#1}}

\def\calE{\mathcal{E}}
\def\bp{{\bf p}}
\def\bk{{\bf k}}
\def\bq{{\bf q}}

\def\br{{\bf r}}

\def\bj{{\bf j}}

\begin{document}

\title{Detecting chiral pairing and topological superfluidity using circular dichroism}
\date{\today}

\author{J.\ M.\ Midtgaard}
\affiliation{ 
Department of Physics and Astronomy, Aarhus University, Ny Munkegade, DK-8000 Aarhus C, Denmark. }
\author{Zhigang \surname{Wu}}
\affiliation{Shenzhen Institute for Quantum Science and Engineering and Department of Physics, Southern University of Science and Technology, Shenzhen 518055, China}
\affiliation{Center for Quantum Computing, Peng Cheng Laboratory, Shenzhen 518055, China}
\author{N.\ Goldman}
\affiliation{Center for Nonlinear Phenomena and Complex Systems,
Universit\'e Libre de Bruxelles, CP 231, Campus Plaine, 1050 Brussels, Belgium}
\author{G.\ M.\ Bruun} 
\affiliation{ 
Department of Physics and Astronomy, Aarhus University, Ny Munkegade, DK-8000 Aarhus C, Denmark. }
\affiliation{Shenzhen Institute for Quantum Science and Engineering and Department of Physics, Southern University of Science and Technology, Shenzhen 518055, China}

\begin{abstract}
Realising and probing topological superfluids is a key goal for fundamental science, with exciting technological promises.  
Here, we show that chiral $p_x+ip_y$ pairing in a two-dimensional topological superfluid can be detected through circular dichroism, namely, as a difference in the excitation rates induced by a clockwise and counter-clockwise circular drive. For weak 
pairing, this difference is to a very good approximation determined by the Chern number of the superfluid, whereas  there is a  non-topological contribution scaling as the superfluid gap squared that becomes signifiant 
 for stronger pairing. This gives rise to  a competition between the experimentally driven goal to maximise 
the critical temperature of the superfluid, and observing a signal given by the underlying topology. Using a combination of strong coupling 
Eliashberg and Berezinskii-Kosterlitz-Thouless  
theory, we analyse this tension for an atomic Bose-Fermi gas, which represents a promising platform for realising a chiral superfluid. 
We identify a wide range of system parameters  
 where both the critical temperature is high and the topological contribution to the dichroic signal is dominant. 
\end{abstract}

\maketitle
\section{Introduction} The realisation and manipulation of topological superfluids and superconductors 
is presently one of the most actively pursued goals  in physics. In addition to 
being interesting from a fundamental science point of view, their Majorana edge modes promise  applications for quantum computing~\cite{Nayak2008}.  
Zero-energy states at the ends of one-dimensional (1D) nanowires have  been observed, consistent with the  presence of Majorana 
modes~\cite{NadjPerge2014,Lutchyn2018}.
So far, there has however been no observation of topological superfluidity in 2D. 
The most promising solid-state candidate  for a 2D topological  superconductor  is  Sr$_2$RuO$_4$, but 
 the precise symmetry of the order parameter in this crystal remains subject to intense debate~\cite{Hicks2010,Mackenzie2017,Kivelson2020}. 
 It has recently been shown that an atomic  2D Fermi gas immersed in a  BEC offers 
a  promising platform for realising  a   topological superfluid~\cite{Wu2016, Midtgaard2016, Midtgaard2017}. 
The fermions form Cooper pairs with chiral symmetry by exchanging 
sound modes in the BEC, and the system offers sufficient flexibility so that one can tune the superfluid critical temperature  to be within experimental reach. 
Experimentally, such a Bose-Fermi mixture has  been realized using $^{173}$Yb-$^7$Li atoms, 
which constitutes an important  step towards the first unequivocal realisation of a  topological  $p_x\!+\!ip_y$ superfluid~\cite{Schaefer2018}.

A key question concerns the detection of topological superfluidity in  atomic gases. Their topological properties  are not easily extracted from 
thermodynamic measurements nor using common probes such as  radio-frequency spectroscopy~\cite{Grosfeld2007}. Contrary to the chiral edge modes of single-particle band structures, which have been detected in experiments~\cite{Cooper2019}, the observation of  Majorana states~\cite{Moeller2011,Gong2012} is complicated by their small number and 
 their particle-hole nature. 

 
 It was recently proposed~\cite{Tran2017,tran2018quantized} and experimentally demonstrated~\cite{Asteria2019} that the topologically invariant Chern number can be detected in  atomic gases through circular dichroism, namely, by analyzing excitation rates upon applying a circular drive. This topological probe was first introduced for non-interacting Chern insulators~\cite{Tran2017}, and later applied to interacting many-body systems~\cite{schuler2017tracing,repellin2019detecting,klein2020stochastic}. Inspired by this approach, we  hereby demonstrate that the chirality of the $p_x\!+\!ip_y$ pairing is revealed in the circular dichroism of the superfluid. 
 For weak pairing, the differential excitation rate obtained from opposite drive orientations, integrated over the drive frequency, is shown to be determined by the Chern number of the topological superfluid, in direct analogy with Chern insulators~\cite{Tran2017}. However, in contrast with the latter case, a non-topological contribution scaling as the superfluid gap squared becomes significant for strong pairing.
The resulting competition  between maximising the superfluid critical temperature 
while detecting a genuine topological signature is analysed for a concrete atomic Bose-Fermi mixture. Using the strong-coupling Eliasberg equations combined with Berezinskii-Kosterlitz-Thouless (BKT) theory, we identify a wide and accessible parameters regime where the superfluid critical temperature is high and the dichroic signal  dominated by the topological Chern number.  Our results demonstrate that the dichroic probe offers an experimentally 
promising pathway to detect topological superfluidity. 

\section{Topological responses in superfluids}
We first establish a connection between circular dichroism, the Hall conductivity, and the Chern number of the superfluid. Consider a 2D system of spin-polarised fermions described by the Hamiltonian 
\begin{gather}
H_0=\int d^2r \, \psi^\dagger({\mathbf r})\left(-\frac {\nabla^2}{2m}\right)\psi({\mathbf r})\nonumber\\+
\frac12\iint d^2rd^2r' \, \psi^\dagger({\mathbf r})\psi^\dagger({\mathbf r}')V({\mathbf r}-{\mathbf r}')\psi({\mathbf r}')\psi({\mathbf r}),\label{eq:ham}
\end{gather}
where $\psi({\mathbf r})$ is the fermion field and $V({\mathbf r}-{\mathbf r}')$ is an interaction giving rise to  pairing, which may result from a $p$-wave
 Feshbach resonance or, as we will consider later, from an induced interaction. Within BCS theory, this $p$-wave superfluid can be described by the Hamiltonian
\begin{align}
H_{\rm BCS} = \sum_\bk \Phi_\bk^\dag H_\bk \Phi_\bk, \quad H_{\mathbf k}={\mathbf h}_{\mathbf k}\cdot \bm{\tau},
\label{HBCS}
\end{align}
where $\Phi_\bk \equiv [a_\bk , a^\dag_{-\bk}]^T$, ${\mathbf h}_{\mathbf k}=(\text{Re}\Delta_{\mathbf k},-\text{Im}\Delta_{\mathbf k},\xi_\bk)^T$ and where $\bm{\tau}=(\tau_1,\tau_2,\tau_3)^T$ with $\tau_i$ the Pauli matrices. Here $\xi_\bk = k^2/2m -\mu$, where $\mu$ is the chemical potential, and $\Delta_\bk$ is the gap parameter ($\hbar\!=\!1$ throughout). The latter 
is taken to have chiral $p$-wave symmetry, i.e.\ $\Delta_{\mathbf k}=\Delta_{k}e^{i\phi}$, where $\phi$ is the polar angle of the momentum ${\mathbf k}$ and $k=|\mathbf k|$,
since this gives the lowest energy for $p$-wave pairing as it have no nodes~\cite{Anderson1961,Lu1991}.
 Indeed, as $\Delta_{k}\propto k$ for $k\ll k_F$ due to  Fermi anti-symmetry, we 
get $\Delta_{\mathbf k}\propto k_x+ik_y$. This results in a topological phase characterised by a Chern number  $C\!=\!-1$ for $\mu>0$ whereas 
 $C\!=\!0$ for $\mu<0$~\cite{Read2000}. The Chern number reads 
\begin{align}
&C=\int\!\frac{d^2k}{4\pi}\frac1{|\mathbf{h_k}|^3}\mathbf{h_k}\cdot\partial_{k_x}{\mathbf h}\times\partial_{k_y}{\mathbf h} \label{ChernFormula}\\
&=\int\!\frac{d^2k}{2\pi}\left[\frac{v_x}{E_{\mathbf k}^3}\text{Im}(\Delta^*_{\mathbf k}\partial_{k_y}\Delta_{\mathbf k})+\frac{\xi_{\mathbf k}}{2E_\k^3}\text{Im}(\partial_{k_x}\Delta_{\mathbf k}\partial_{k_y}\Delta^*_{\mathbf k})\right] ,\notag
\end{align}
where $v_x=k_x/m$ and $E_{\mathbf k}=\sqrt{\xi_{\mathbf k}^2+|\Delta_{\mathbf k}|^2}$ is the BCS quasiparticle energy. 
The second term in the integrand scales as $\Delta_{\mathbf k}^2/\mu^2$ so that the Chern number can be approximated by the first term in the regime $\Delta_{\mathbf k}\ll\mu$.

We now show that the Chern number in Eq.~\eqref{ChernFormula} can be extracted from circular dichroism, namely, by monitoring excitation rates upon a circular drive~\cite{Tran2017,Asteria2019}. We consider a circular drive of the form 
\begin{align}
V_\pm(\r;\q) &= 2 \mathcal E \left(x\cos\Omega t \pm  y \sin\Omega t \right)\cos{ \bq \cdot \br},
 \label{perturbation}
\end{align}
and we will set $\bq\!\rightarrow0$ at the end of calculations, corresponding to a uniform circular shaking~\cite{Asteria2019}. This reads
\begin{align}
V_\pm(\q)=  \frac {\mathcal{E}} {i}\left[ -\frac{\partial \tilde n(\bq)}{\partial q_x}  \pm i \frac{\partial \tilde n(\bq)}{\partial q_y}\right ]e^{-i\Omega t} + \text{h.c.}, 
\end{align}
in second quantization, where $n(\r)=\psi^\dagger({\mathbf r})\psi({\mathbf r})$ is the density operator, $n(\bq)$ is its Fourier transform and $\tilde n(\bq) = [n(\bq)-n(-\bq)]/2$. 
Within linear response, the excitation rate out of the ground state of $H_0$ induced by  $ V_\pm(\q) $ 
can be calculated using Fermi's golden rule as 
\begin{align}
\Gamma_{\pm}(\q,\Omega) = & 2\pi {\mathcal{E}^2} \sum_f \left | \left \langle f \left | \frac{\partial \tilde n(\bq)}{\partial q_x}  \pm i \frac{\partial \tilde n(\bq)}{\partial q_y} \right |g \right \rangle \right |^2  \nn\\
&\times \delta(E_f-E_g- \Omega),
\label{Gammapm}
\end{align}
where $|g\rangle$ and $|f\rangle$ denote the ground and excited states of $H_0$ with energy $E_g$ and $E_f$, respectively. 

The observable of interest is provided by the differential integrated rate (DIR), which is defined as~\cite{Tran2017} 
\begin{align}
\Delta\Gamma=\lim_{\bq\rightarrow 0}\frac12\int_0^\infty\!d\Omega[\Gamma_{+}(\q,\Omega)-\Gamma_{-}(\q,\Omega)]. 
\label{HeatingRateDef}
\end{align}
Substituting Eq.~(\ref{Gammapm}) into Eq.~(\ref{HeatingRateDef}), we find 
\begin{align}
\Delta\Gamma&=-\pi i \mathcal{E}^2\lim_{\bq\rightarrow 0} \left \langle g \left |\left [\frac{\partial  n(\bq)}{\partial q_x}, \frac{\partial  n(-\bq)}{\partial q_y}\right ]\right |g\right \rangle,
\label{DGamman}
\end{align}
where we have used momentum conservation to eliminate terms. 
We now use the continuity equation to write $\Delta \Gamma$ in terms of the density-current correlation function. From 
$\partial_t n(\br,t) + \nabla\cdot \bj(\br,t) = 0$, we find
\begin{align}
\langle g|  n(\bq) |f\rangle = \frac{\langle g| \bq\cdot \bj(\bq) |f\rangle }{E_f -E_g},
\label{iden}
\end{align}
where the Fourier transform of the current reads 
\begin{align}
\bj(\bq)\!=\!(1/2mi)\int d^2r e^{ - i \bq \cdot \br}  [\psi^\dagger({\mathbf r})\nabla \psi({\mathbf r})\!-\!\text{h.c.}].
\end{align}
Using Eqs.~\eqref{DGamman}-\eqref{iden} and noting that 
\begin{align}
\lim_{q_y\rightarrow 0}\lim_{q_x\rightarrow 0} \langle g|{\partial_{q_y} n(\bq)}|f\rangle= \lim_{q_y\rightarrow 0}\lim_{q_x\rightarrow 0}  \langle g|{ n(\bq)}/{ q_y}|f\rangle ,
 \end{align}
we find the relation
\begin{align}
\Delta \Gamma / \mathcal A = 2 \pi \mathcal{E}^2\sigma_{xy} ,
\label{DGammansig}
\end{align}
which connects the DIR to the static Hall conductivity
\begin{align}
\sigma_{xy}&\equiv \lim_{q_y\rightarrow 0}\lim_{q_x\rightarrow 0}\lim_{\omega\rightarrow 0 } \frac{1}{i\mathcal A\omega}\chi_{j_x,j_y}(\bq,\omega)\notag\\
&=\lim_{q_y\rightarrow 0}\lim_{q_x\rightarrow 0}\lim_{\omega\rightarrow 0 } \frac{1}{i \mathcal Aq_y}{\chi_{j_x,n}(\bq,\omega) }.
\label{Hallc2}
\end{align}
Here $\mathcal A$ is the  system's area, and $\chi_{A,B}(\q,\omega)$ is the Fourier transform of the retarded correlation function 
  \begin{align}
\chi_{A,B}(\q,t-t') = -i \theta(t-t') \braket{[A(\q,t),  B(-\q,t')]} , 
\end{align}
with $\theta(x)$ the Heaviside function. We note that the specific order of limits (taking $\omega\!\rightarrow\!0$ before $\bq\!\rightarrow\!0$) is crucial, since the more standard order $\lim_{\omega\rightarrow 0}\lim_{\bq\rightarrow 0 } \frac{1}{i\omega}\chi_{j_x,j_y}(\bq,\omega)$ yields zero for a translationally invariant system~\cite{Giuliani2005}; this subtlety also arises when analysing edge currents~\cite{Volovik1992,Volovik1988,Gorya1998,Stone2004}. 

Besides,   Eq.~\eqref{DGammansig} was obtained by taking the finite nature of realistic systems into account. In particular,
one would obtain an additional factor of $1/2$ for a strictly translationally invariant  system. Indeed, when deriving Eq.~\eqref{DGamman}, we use that 
$\langle A({\mathbf q}),B(\mathbf q')\rangle\propto\delta_{\mathbf q,-\mathbf q'}$ for a strictly infinite  translationally invariant system. From this, it follows 
that $\lim_{\mathbf q\rightarrow 0}\langle n(\mathbf q)n(\mathbf q)\rangle=0$ 
and such terms can be discarded. However, for a finite physical system of size $L$, momentum is only defined with a resolution $\sim 1/L$.
This means that $\langle n(\mathbf q)n(\mathbf q)\rangle$ starts to become non-zero for $q\lesssim1/L$ and in particular
$\lim_{\mathbf q\rightarrow 0}\langle n(\mathbf q)n(\mathbf q)\rangle= \langle n(\mathbf 0)n(\mathbf 0)\rangle$ for a finite system, which leads to the extra factor of 
$2$ on the right hand side of Eq.~\eqref{DGammansig}.
 Physically, it means that a finite system cannot distinguish between a force with wave length much greater than the system size from a uniform force. We note that uniform circular shaking ($\mathbf q\!=\!0$) can be realized in ultracold-atom experiments.
We also point out that Eq.~\eqref{DGammansig}  is universal:~it can also be derived from Kramers-Kronig relations~\cite{Tran2017,repellin2019detecting,bennett1965faraday,souza2008dichroic}, noting that the excitation rates $\Gamma_{\pm}(\Omega)$ are related to the power absorbed upon the circular drive $P_{\pm}(\Omega)=\Omega \Gamma_{\pm}(\Omega)$.

The Hall conductivity of a superfluid was previously shown to be related to a Chern number; see Refs.~\cite{Volovik1992,Volovik1988} in the context of chiral $^3$He superfluids. 
Using BCS theory, the current-density correlation function can be written as~\cite{Mahan2000book} 
\begin{equation}
\chi_{j_x,n}(q) = -\sum_k \frac{k_x}{2m} \text{tr}\left[\mathcal{G}_0(k-q/2)\mathcal{G}_0(k+q/2)\tau_3\right]
\label{eq:CurrentDensity}
\end{equation}
where in shorthand notation $k = (\vec{k}, \omega_n)$ with $\omega_{n}$ a fermionic Matsubara frequency, and $\mathcal{G}_0$ is the BCS Green's
function. We have 
\begin{equation}
\mathcal{G}_0(k) = \int_{-\infty}^{\infty} \frac{d\omega}{(-\pi)} \frac{\Im \mathcal{G}_0(\k, \omega)}{i\omega_n - \omega}
\end{equation}
where
\begin{align}
\Im \mathcal{G}_0(\k, \omega) =& -\frac{\pi}{2E_\k}
\begin{pmatrix}
\omega + \xi_\k & -\Delta_\k \\
-\Delta_\k^\ast & \omega - \xi_\k
\end{pmatrix}\nonumber\\
&\times
\left[ \delta(\omega - E_\k) - \delta(\omega + E_\k) \right].
\end{align}
Inserting this in Eq.~\eqref{eq:CurrentDensity}  and performing the Matsubara sum yields to first order in  $\vec{q}$
\begin{align}
\lim_{\omega \to 0} \chi_{j_x,n}(\vec{q},\omega) =  &-\frac{i q_x}{\mathcal{V}} \sum_\k \frac{k_x}{2m} \frac{ \Im\Delta_\k \partial_{k_x} \Delta_\k^\ast }{E_\k^3} \nonumber\\
&- \frac{i q_y}{2\mathcal{V}} \sum_\k \frac{k_x}{2m} \frac{ \Im\Delta_\k \partial_{k_y} \Delta_\k^\ast }{E_\k^3}.
\end{align}
The first of these terms vanish, since the summand is odd in $k_y$. This can be seen if we fix the phase of the gap function and look at, for instance, the simple example $\Delta_\k = k_x + ik_y$. With this, it is clear that
\begin{align}
\lim_{{\mathbf q} \rightarrow0} \lim_{\omega \to 0} \frac{\chi_{j_x,n}({\mathbf{q}},\omega)}{iq_y}&=\int\!\frac{d^2\vec{k}}{8\pi^2}\frac{k_x}m\frac{\text{Im}[\Delta_\mathbf{k}^*\partial_{k_y}\Delta_\mathbf{k}]}{E_\mathbf{k}^3}\nonumber\\
&=\frac{C}{4\pi}+\mathcal{O}(\Delta^2/\mu^2),
\label{ChernCurrentDensity}
\end{align}
where the last equality is obtained by comparing with the Chern number in Eq.~\eqref{ChernFormula}. In contrast with the case of Chern insulators, where
the Hall conductivity is genuinely topological in the thermodynamic limit~\cite{Thouless1982}, the Hall response of the superfluid [Eq.~\eqref{ChernCurrentDensity}] contains a correction scaling as $\mathcal{O}(\Delta^2/\mu^2)$. This result was previously related to the fact that the edge current of a chiral $p$-wave superconductor is not strictly topological, as opposed to the  
presence of edge (Majorana) states~\cite{Taylor2012,Huang2014,Huang2015}.

Finally, combining Eqs.~\eqref{DGammansig} and \eqref{ChernCurrentDensity} yields the central result of this work,
 \begin{align}
  \Delta\Gamma / \mathcal A= (1/2) \, \mathcal{E}^2C+\mathcal{O}(\Delta^2/\mu^2),
\label{HeatingRatChern}
\end{align}
which shows that the DIR related to the dichroic probe is closely related to the 
Chern number of the superfluid phase:~this observable  exhibits a jump proportional to the Chern number to order $\mathcal{O}(\Delta^2/\mu^2)$ whenever the superfluid enters the topological phase with $C=-1$.

    
\section{Dichroic probe for a topological Bose-Fermi mixture}
We now explore the dichroic probe for a concrete system consisting of a 2D gas of fermionic atoms  
immersed in a 3D  BEC. The  fermions interact by exchanging sound modes in the BEC, which leads to an induced attractive interaction and Cooper 
pairing~\cite{Wu2016}.
 Since both the range and  strength of the induced interaction can be varied, one 
can tune the mixture in order to  reach a high critical temperature. This makes such a mixture a strong candidate for observing a  chiral  pairing. Recently,  progress towards realising this goal was reported 
with the experimental realisation of  a  $^{173}$Yb-$^7$Li mixture~\cite{Schaefer2018}. 
We now analyse how the dichroic probe can be used to detect  topological  pairing in this specific system. 

Due to the finite speed of sound in the BEC, the   interaction between the fermions mediated by the bosons is not instantaneous, thus giving rise to retardation effects. The latter are included in the frequency-dependent Eliashberg 
equations as explained in the appendix. It has  been shown that retardation effects are small when the bosons in the BEC are light compared to the fermions
such as for the  $^{173}$Yb-$^7$Li mixture~\cite{Kinnunen2018}. 
The induced interaction is then close to the static Yukawa form 
\begin{equation}
V(\mathbf r) = -\frac{a_{\rm eff}^2n_B m_B}{\pi} \frac{\exp(-\sqrt{2}r/\xi_B)}{r}.
\label{Interaction}
\end{equation} 
Here, $n_B$ and $m_B$ is the density and mass of the bosons, $\xi_B=1/\sqrt{8\pi n_Ba_B}$ is the BEC healing length  with $a_B$ the boson-boson scattering length,
 and $a_{\rm eff}$ is the  mixed dimensional Bose-Fermi scattering length. 

According to Eq.~\eqref{HeatingRateDef}, one should measure the differential rate $\Gamma_+-\Gamma_-$ integrated over all frequencies. However, any real 
measurement necessarily 
introduces an upper cut-off frequency $\Omega_c$ above which there is no signal~\cite{Asteria2019}. Using Eqs.~\eqref{DGammansig} and \eqref{ChernCurrentDensity}, 
 the resulting signal reads
 \begin{equation}
\frac{\Delta \Gamma_\text{trunc} (\Omega_c)}{\mathcal A\mathcal{E}^2}\!\equiv\!\int\!\frac{d^2k}{4\pi}\frac{k_x}m\frac{\text{Im}[\Delta_\mathbf{k}^*\partial_{k_y}\Delta_\mathbf{k}]}{E_\mathbf{k}^3} 
\theta(\Omega_c - 2E_\vec{k})
\label{HeatingRateTrunc}
\end{equation}
  The cut-off $\theta(\Omega_c - 2E_\vec{k})$ 
  reflects that the probe breaks 
 pairs with energy  $2E_\vec{k}$ in the long wave length limit. We note that $\Delta\Gamma\!=\!\lim_{\Omega_c \to \infty}\Delta\Gamma_\text{trunc}(\Omega_c)$.

\begin{figure}[htb]
\centering
\includegraphics[width=\columnwidth]{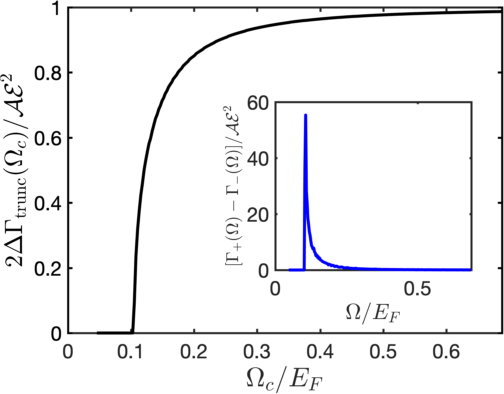}
\caption{(Color online).  The differential integrated rate $\Delta \Gamma_\text{trunc} (\Omega)$   between a clockwise and counterclockwise perturbation as a 
function of the cut-off frequency. The insert shows the difference in the heating rates between a clockwise and counterclockwise perturbation as a function of frequency. }
\label{fig:heatingrate}
\end{figure}

 In Fig.~\ref{fig:heatingrate}, we plot $\Delta \Gamma_\text{trunc} (\Omega_c)$ for a $^7$Li-$^{173}$Yb mixture with a 
 Bose-Fermi coupling $n_B^{1/3} a_{\rm eff} = 0.12$, BEC gas parameter $n_B^{1/3} a_B = 0.1$,
  and density ratio $n_F^{1/2}/n_B^{1/3} = 0.5$, where $n_F$ is the 2D Fermi density.
 These results are obtained by first solving the BCS equations self-consistently at zero temperature  and 
 then evaluating the DIR from Eq.~\eqref{HeatingRateTrunc}. The numerical solution indeed confirms the $p$-wave form of pairing 
 $\Delta_{\bf k} = \Delta_k e^{i\phi}$ where $\Delta_k \propto k$ for small momenta.
The DIR is zero for cut-off frequencies below twice the gap, i.e.\ for $\Omega_c \lesssim 0.1 E_F$
where $E_F$ is the 2D Fermi energy, reflecting that  there is  
not enough energy in the probe to break pairs. Above this threshold, the DIR quickly 
 converges towards to the Chern 
number for $\Omega_c \gtrsim E_F$.  Since $\Delta \simeq 0.05E_F \ll E_F$ for this set of parameters, the deviation 
of  $2\Delta\Gamma/\mathcal{AE}^2$ away from the Chern number is small. 
We also plot in Fig.~\ref{fig:heatingrate} the differential rate at a given frequency $\Omega$, 
\begin{align}
\frac12[\Gamma_+(\Omega)-\Gamma_-(\Omega) ]=\frac{\partial}{\partial \Omega }\Delta\Gamma_\text{trunc}(\Omega).
\end{align}
This difference is  large for frequencies  
 just above the threshold given by twice the gap, where the density of states of the superfluid is  highest,  
  and Fig.~\ref{fig:heatingrate}  shows that that one only needs to measure  the difference up to a few times the pairing gap  to resolve the Chern number. 

\begin{figure}[htb!]
\centering
\includegraphics[width=\columnwidth]{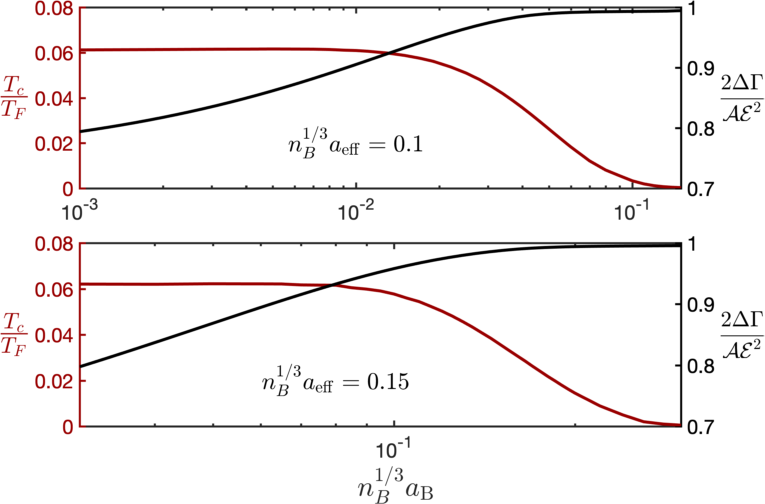}
\caption{(Color online). The critical temperature (red) and the differential integrated rate $\Delta \Gamma$ (black) as a function of the BEC gas parameter $n_B^{1/3}a_\text{B}$ for two different Bose-Fermi interaction strengths. }
\label{fig:chern}
\end{figure}

One of the appealing features of the Bose-Fermi mixture is that the critical temperature for the 2D  superfluid can be tuned to be close to the maximum 
value $T_c/T_F=1/16$ allowed by BKT theory. Maximising $T_c$ will however also increase the gap and thereby increase  corrections 
to the DIR away from the Chern number  as seen from Eq.~\eqref{HeatingRatChern}.  
To analyse this tension, we plot in Fig.~\ref{fig:chern} the DIR $\Delta \Gamma$ at zero temperature  and the critical temperature $T_c$ as a 
function of the gas parameter $n_B^{1/3} a_\text{\rm B}$ for $n_F^{1/2}/n_B^{1/3} = 0.5$ and two different 
Bose-Fermi interaction strengths. The critical temperature is calculated by combining  strong coupling 
Eliashberg and BKT theory, which includes the frequency dependence of the gap; 
see Refs.~\cite{Wu2016} and the appendix for details. We see that the critical temperature increases with decreasing gas parameter reflecting that 
 the range of interaction in Eq.~\eqref{Interaction}, given by the  BEC coherence length, increases. The gap consequently also increases  
 leading to a larger correction term for the DIR away from $\Delta \Gamma\!=\!\mathcal{AE}^2{\mathcal C}/2$.
 Nevertheless, Fig.~\ref{fig:chern} shows that there is 
a significant region where both the DIR is close to the topological value and the critical temperature is close 
to its maximum value $T_c/T_F=1/16$. Note that  we expect our calculation to give a lower bound on the DIR, since BCS theory likely overestimates the gap. 

\begin{figure}[htb]
\centering
\includegraphics[width=\columnwidth]{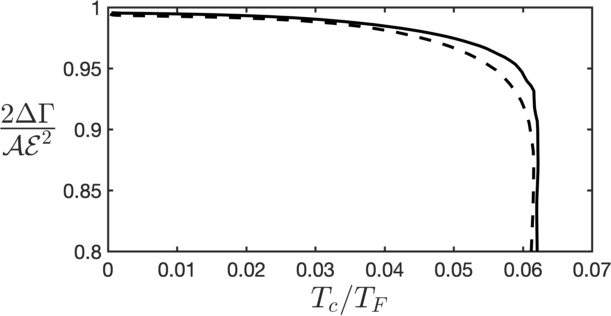}
\caption{The DIR as a function of the critical temperature for the same parameters as in Fig.~\ref{fig:chern}. The dashed line corresponds to  $n_B^{1/3}  a_\text{eff}= 0.1$ and the solid line  to $n_B^{1/3}  a_\text{eff}= 0.15$.}
\label{fig:tg}
\end{figure}

To further illustrate the competition between maximising the critical temperature and measuring a value of $\Delta\Gamma$ determined by the underlying topology,  
we plot $\Delta\Gamma$ at zero temperature as a function of $T_c$ in Fig.~\ref{fig:tg} for the same parameters as in Fig.~\ref{fig:chern}.  
This demonstrates that in order for the dichroic probe to yield a value close to that given by the Chern number, one should cool to around $T\sim 0.06E_F$.
Since temperatures down to $T\simeq 0.03E_F$ have been obtained for 2D Fermi gases~\cite{Luick2019,Sobirey2020,Ries2015}, this is within present day technology making 
our scheme promising for detecting topological superfluidity.
 It  also shows that 
  a stronger Bose-Fermi  interaction strength is slightly more favorable although the difference between the two interaction strengths is   small.

For $T=0$,
BCS theory has been shown to be  surprisingly accurate  even for strong coupling where the Cooper pairs are tightly bound and the system 
is in  the so-called BEC regime~\cite{Pieri2005}. It follows that our calculation of the DIR is reliable even in this regime, where the correction term $\mathcal O(\Delta^2/\mu^2)$
away from the quantized value is large. Any non-zero value however indicates chiral pairing, since the DIR is zero in a phase with time-reversal symmetry.
Our scheme thus provides a way to observe the topological phase transition to a trivial phase when  $\mu$ becomes negative deep in the  BEC regime. 


\section{Conclusion}
We showed that chiral $p_x+ip_y$ pairing in a 2D superfluid can be detected through circular dichroism. Contrary to the case of topological insulators~\cite{Tran2017}, the DIR is not purely dictated by the Chern number due to a correction term scaling 
 as $\Delta^2/E_F^2$, giving rise to a competition between maximising the critical temperature of the superfluid and 
 observing   the  Chern number from such a dichroic probe. As a concrete example, 
  we  considered  an atomic Bose-Fermi mixture. Using a combination of Eliashberg and BKT theory, it was 
  demonstrated that there is in fact a wide range of values for the system parameters where both the critical 
  temperature is high and the dichroic signal close to the value given by the Chern number. This combined with the fact  that a
   similar scheme was recently successfully applied to detect topological order in a Chern Bloch band~\cite{Asteria2019}, leads to the  conclusion that the dichroic probe is a 
  strong candidate for  detecting topological $p_x+ip_y$ pairing in an atomic system. 
 
{\it Acknowledgements.---} J.M.M.\ and G.M.B.\ wish to acknowledge the support of  the Danish Council of Independent Research --
 Natural Sciences via Grant No. DFF - 4002-00336. Z.W.
acknowledges the support by the National Science Foundation of China (Grant No.11904417)
and the Key-Area Research and Development Program of GuangDong Province (Grant No.2019B030330001). N.G. is supported by the ERC Starting Grant TopoCold, and the Fonds De La Recherche Scientifique (FRS-FNRS, Belgium).


%


\appendix



\section{Calculation of the superfluid transition temperature}
\label{ap:Tc}
Here, we outline the calculation of the superfluid transition temperature for the 2D $^{173}$Yb gas immersed in a 3D $^7$Li BEC. First, we solve the following frequency-dependent gap equation at a finite temperature $T$~\cite{Kinnunen2018}
\begin{align}
\Delta(\bp,i\omega_n) =& -T \sum_{m} \int\frac{d\bq}{(2\pi)^2}V_{\rm ind}(\bp-\bq,i\omega_n-i\omega_m)\nonumber\\ 
&\times\frac{\Delta(\bq,i\omega_m) }{\omega_m^2+\calE^2(\bq,i\omega_m)},
\label{gapeq_fre}
\end{align}
where $\calE(\bq,i\omega_m) = \sqrt{\xi_\bq^2 + |\Delta(\bq,i\omega_m)|^2}$. Here the frequency-dependent induced interaction $ V_{\rm ind}(\bq,i\omega_\nu)$ is given by 
\begin{align}
 V_{\rm ind}(\bq,i\omega_\nu) = -n_Bm_Bg^2\left [\left(\frac{1}{\kappa_+}+\frac{1}{\kappa_-}\right )\right.\nonumber\\
\left.  + \frac{1}{\sqrt{1-(\omega_\nu/g_Bn_B)^2}}\left(\frac{1}{\kappa_+}-\frac{1}{\kappa_-}\right ) \right ], 
 \label{Vindf}
 \end{align}
 where $ \kappa_\pm = \sqrt{2m_Bg_Bn_B\left[1\pm \sqrt{1- (\omega_\nu/g_Bn_B)^2}\right]+\bq^2}$. Along with a number equation, this constitutes the Eliashberg equations of the superfluid~\cite{Mahan2000book}.
 
Since the Fermi system is 2D, the superfluid transition is driven by vortex-antivortex proliferation and the critical temperature $T_{\rm BKT}$ is determined by the  Kosterlitz-Thouless condition~\cite{Wu2016}
\begin{align}
T_{\rm BKT} = \frac{\pi}{8m_F^2}\rho_s\left (\left \{\Delta(i\omega_n)\right \}, T_{\rm BKT}\right ).
\label{TBKT}
\end{align}
Here,  $\rho_s$ is the superfluid mass density and is a function of the gap parameters and the temperature. Neglecting the renormalisation of the 
 interaction between vortex pairs, $\rho_s$  can be estimated as 
\begin{align}
\rho_s = \rho_0 + \frac{T}{2}\sum_{n}\int \frac{d\bp}{(2\pi)^2} p^2 \frac{ \calE^2(\bp,i\omega_n)-\omega_n^2}{\left[\omega_n^2+\calE^2(\bp,i\omega_n)\right ]^2},
\label{ns}
\end{align} 
where $\rho_0 = m_F n_F$. Solving Eq.~(\ref{TBKT}) self-consistently using Eqs.~(\ref{ns}) and the frequency-dependent gap parameters obtained from Eq.~(\ref{gapeq_fre}), we obtain the superfluid transition temperatures shown in the main text.

\end{document}